\documentclass[aps,prd,onecolumn,preprintnumbers,groupedaddress,showpacs,nofootinbib,amssymb]{revtex4}
\usepackage{graphicx,color}
\usepackage{amsmath}
\usepackage{amssymb}
\usepackage{amsfonts}
\usepackage{bm}
\usepackage{cancel}
\usepackage{graphicx}
\usepackage{epsf}
\usepackage{bm}
\usepackage{amsmath}
\usepackage{amsfonts}
\usepackage{amssymb}
\usepackage{color}
\usepackage{epstopdf}

\allowdisplaybreaks[4]

\newcommand{\bea}{\begin{eqnarray}}
\newcommand{\eea}{\end{eqnarray}}
\newcommand{\nn}{\nonumber \\}
\newcommand{\e}{\mathrm{e}}

\newcommand{\ba}{\begin{eqnarray}}
\newcommand{\ea}{\end{eqnarray}}
\newcommand{\be}{\begin{equation}}
\newcommand{\ee}{\end{equation}}

\begin{document}

\title{The role of energy conditions in $f(R)$ cosmology}
\author{S. Capozziello$^{1,2,3}$, S. Nojiri$^{4,5,6}$, S. D. Odintsov$^{7,8}$}

\affiliation{\it $^1$ Dipartimento
di Fisica``E. Pancini", Universit\`{a} di Napoli {}``Federico II'' \\
$^2$INFN Sez. di Napoli, Compl. Univ. di Monte S. Angelo, Edificio G, Via 
Cinthia, I-80126, Napoli, Italy \\
$^3$Gran Sasso Science Institute, Via F. Crispi 7, I-67100, L' Aquila, Italy \\
$^4$ Department of Physics,
Nagoya University, Nagoya 464-8602, Japan \\
$^5$ Kobayashi-Maskawa Institute for the Origin of Particles and the Universe, 
Nagoya University, Nagoya 464-8602, Japan \\
$^6$ KEK Theory Center, High Energy Accelerator Research Organization (KEK),
Oho 1-1, Tsukuba, Ibaraki 305-0801, Japan \\
$^7$ICREA, Passeig Luis Companys,
23, 08010 Barcelona, Spain, \\
$^8$ Institute of Space Sciences (IEEC-CSIC)
C. Can Magrans s/n, 08193 Barcelona, Spain}

\date{\today}
\begin{abstract}
Energy conditions can play an important role in defining the cosmological 
evolution. Specifically acceleration/deceleration
of cosmic fluid, as well as the emergence of Big Rip singularities, can be related 
to the constraints imposed by the energy 
conditions. Here we discuss this issue for
$f(R)$ gravity considering also conformal transformations. 
Cosmological solutions and equations of state can be
classified according to energy conditions. The qualitative change of 
some energy conditions for transformations from the Jordan frame to the Einstein 
frame  is also observed.
\end{abstract}


\maketitle

\section{Introduction \label{SecI}}

The observed cosmic acceleration \cite{Riess:1998cb,Riess:1998dv,Perlmutter:1998np,Spergel:2003cb,Eisenstein:2005su} 
points out that a revision of the 
 cosmological picture, based on the General Relativity (GR) and the 
standard model of particles, is needed. The puzzle can be addressed either 
introducing some form of dark energy or assuming 
modifications of GR. In other words, one can act either on the r.h.s. of the Einstein equations 
by introducing some new matter-energy fluid on the l.h.s. modifying or improving geometry. 
In this latter perspective, $f(R)$ gravity is the straightforward modification of GR where, instead 
of assuming the gravitational action strictly linear in the Ricci
scalar $R$, one takes into account a 
general function of $R$. The paradigm is that the form of $f(R)$ can be fixed according to 
the cosmological and astrophysical observations ranging from local to cosmological scales
\cite{Nojiri:2010wj,Capozziello:2011et,Capozziello:2012ie,DeMartino:2013zua,Olmo:2011uz,
Nojiri:2017ncd,Koivisto:2006ie,Borowiec:2006hk,Li:2006vi,Movahed:2007cs}.

Beside phenomenological approaches, first principles like energy conditions, causal structure and 
the classification of singularities can be considered to restrict the possible forms of $f(R)$ 
models \cite{Clifton:2005at,Reboucas:2009yw,Santos:2010tw, Santos:2007bs,Atazadeh:2008mh,Wang:2012mws,Wang:2012rw,Banijamali:2011up}. 
In particular, energy conditions, originally formulated in Ref. \cite{Hawking:1973uf} for GR, can play 
an important role to fix physically consistent $f(R)$ models \cite{Santos:2007bs}. 
In this debate, the role of conformal transformations is crucial because, also if the Jordan and Einstein 
frames are mathematically equivalent, the meaning of energy conditions can depend on the frame where 
they are formulated \cite{Barrow:1988xh,Magnano:1993bd,Capozziello:2003gx,Olmo:2006eh}. 
In particular, the effective pressure and effective energy definitions changes according to the frame
\cite{PerezBergliaffa:2006ni,Santos:2007bs,Atazadeh:2008mh,
Wang:2012mws,Wang:2012rw,Banijamali:2011up,Kung:1995he,Kung:1995nh} 
not only in $f(R)$ gravity but also 
in other alternative theories of gravity~\cite{Garcia:2010xz}. 
In general, it is important to define the role of further geometrical terms in the stress-energy tensor 
\cite{Capozziello:2012uv,Albareti:2012va,Capozziello:2014bqa,Capozziello:2013vna} and to recast 
the energy conditions accordingly. Conformal transformations and their physical meaning are crucial 
in the perspective of determining self-consistent energy conditions. For review, see 
\cite{Maeda:1988rb,Cotsakis:1993vm,Teyssandier:1995wr,Schmidt:1995ws,Cotsakis:1995wt,
Capozziello:1996xg,Tsamparlis:2013aza,Quiros:2011wb,Chatterjee:2012zh}.

In this paper, we are considering the role of energy conditions in 
of $f(R)$ cosmology. In particular, we discuss the conformal transformations of the $f(R)$ 
effective energy-momentum tensor. This issue is extremely relevant to 
address the attractive/repulsive behavior
of $f(R)$ cosmological models in relation to the equation of state.

The paper is organized as follows.
In Sec.~\ref{SecII}, we consider the energy conditions in GR. Their 
definition for Extended Theories of
Gravity (ETG) is taken into account in Sec.~\ref{SecIII}. 
The effective energy-momentum tensor, containing curvature terms, is discussed 
in Sec.~\ref{SecIV}.
The relations of this generalized energy-momentum tensor to the cosmological 
equation of state are
considered in Sec.~\ref{SecV}. As an example of the above general results, we 
assume the case of power-law $f(R)$ gravity
in Sec.~\ref{SecVI}. Conclusions are drawn in Sec.~\ref{SecVII}.

\section{Energy conditions in General Relativity \label{SecII}}

Let us start from the Einstein field equations 
\begin{equation}
\label{geometric}
\left(R_{\mu\nu}-\frac{1}{2}g_{\mu\nu}R\right)=\frac{\kappa^2}{2}T_{\mu\nu},
\end{equation}
where $R_{\mu\nu}$ is the Ricci tensor, $R$ is the Ricci scalar, and 
$T_{\mu\nu}$
is energy-momentum tensor of the matter fields. Such equations determine the causal and 
geodesic structure of space-time. 
 The Einsten field equations can be written also as
\begin{equation}
\label{material}
R_{\mu\nu}=\frac{\kappa^2}{2}\left(T_{\mu\nu}-\frac{1}{2}Tg_{\mu\nu} \right)\,,
\end{equation}
where the analog role of matter and geometry
into dynamics is evident. Due to this aspect, we can deal with 
\textit{geometrodynamics}
after Wheeler \cite{Misner:1974qy}.
Since such equations are addressing the causal (metric) and geodesic structure of the space-time, the 
energy-momentum tensor has to satisfy some conditions. 
We can take into account a timelike vector $u^\alpha$ normalized 
as $u^\alpha u_{\alpha}=-1$ for the signature $(-+++)$. It is
the four-velocity of an observer in space-time, and an arbitrary, 
future-directed null vector $k^\alpha$, i.e. $k^\alpha k_{\alpha}=0$.
The energy conditions are contractions of timelike or null vector fields 
with respect to the Einstein tensor and
energy-momentum tensor coming from field Eqs. (\ref{geometric}) or (\ref{material}). 
We obtain four conditions \cite{Hawking:1973uf,Poisson} which are
\begin{itemize}
\item The {\bf WEC} (WEC) which states that at each point of the 
space-time $\mbox{p}\in \mathcal{M}$
the energy-momentum tensor satisfies the inequality
\begin{equation}
\label{wec}
T_{\mu\nu}u^\alpha u^\beta\geq0\, .
\end{equation}
for any timelike vector $u\in T_{\mbox{p}}\mathcal{M}$. If $u^\alpha$ is a 
four-velocity of an observer, then the
quantity $T_{\mu\nu}u^\alpha u^\beta$ is the local energy density 
and the inequality (\ref{wec})
is equivalent to the assumption that the energy density of a given matter source, 
measured by an arbitrary observer, is non-negative. The canonical form of the 
energy-momentum tensor \cite{Hawking:1973uf} can be written
in the orthonormal basis as $T^{\mu\nu}=\mathrm{diag}(\rho,p_1,p_2,p_3)$ and then, one obtains 
\begin{equation}
\label{wec2}
\rho\geq0\, ,\quad \rho +p_i>0\, ,\quad i=1,2,3\, .
\end{equation}
Following \cite{Capozziello:2014bqa}, it can be written as
\begin{equation}
R_{\mu\nu}u^\mu u^\nu \geq -\frac{\kappa^2}{4}(\rho-\sum^3_{i=1} p_i)\, .
\end{equation}
\item The {\bf Null Energy Condition} (NEC) considers future-directed
null vector $k^\mu$
\begin{equation}
T_{\mu\nu}k^\alpha k^\beta\geq0\, ,
\end{equation}
from which one gets $\rho+p_i\geq0$.
\item The {\bf Dominant Energy Condition} (DEC) staes that matter flows 
along timelike or null world lines.
By contracting the energy-momentum tensor with an arbitrary, future-directed, 
timelike vector fields, the
quantity $-T^\mu_{\,\nu}u^\nu$ becomes a future-directed, timelike or null vector 
field. It is called the matter momentum
density that a given observer can measure. This means that, in any orthonormal basis, 
the energy dominates the other components
of the energy-momentum tensor being $T^{00}\geq|T^{ij}|$:
\begin{equation}
\rho\geq0\, ,\quad \rho\geq |p_i|\, .
\end{equation}
\item The {\bf Strong Energy Condition} (SEC)
\begin{equation}
\left( T_{\mu\nu} -\frac{1}{2}Tg_{\mu\nu} \right)u^\mu u^\nu\geq 0
\end{equation}
is a statement about the Ricci tensor:
\begin{equation}
R_{\mu\nu}u^\mu u^\nu\geq0\, ,
\end{equation}
and together with the Raychaudhuri equation 
\cite{Raychaudhuri:1953yv,Ehlers:2006aa,GFREllisPramana,Kar:2006ms}
gives that gravity has to be attractive.
\end{itemize}
All these considerations are related to standard matter which satisfies regular 
equations of state and is minimally coupled to the geometry. They can be generalized to other theories 
of gravity assuming that at least causal structure is preserved.

\section{Energy conditions in Extended Theories of Gravity \label{SecIII}}

Any alternative theory of gravity should be confronted with energy conditions which assign the 
fundamental causal and geodesic structure of space-time. In particular Extended Theories of 
Gravity (ETGs) \cite{Nojiri:2010wj,Capozziello:2011et,Capozziello:2012ie}, which are straightforward 
extensions of the Einstein gravity, can be recast in such a way to be dealt under the standard of energy 
conditions. As discussed in 
\cite{Capozziello:2014bqa,Capozziello:2013vna}, the field equations of any ETG can be written in the 
form 
\begin{equation}
g(\Psi^i)(G_{\mu\nu}+H_{\mu\nu})=\frac{\kappa^2}{2}T_{\mu\nu}\, ,
\end{equation}
where $G_{\mu\nu}=R_{\mu\nu}-\frac{1}{2}g_{\mu\nu}R$ is the Einstein tensor, 
$g(\Psi^i)$ is a generalized
 coupling with the matter fields which contributes to the energy-momentum 
tensor $T_{\mu\nu}$. $\Psi^i$ 
represents curvature invariants and/or gravitational fields which contributes to 
the dynamics. 
$H_{\mu\nu}$ is a geometric tensor term including
all geometrical modifications given by the given ETG. 
General Relativity is recovered assuming
$g(\Psi^i)=1$ and $H_{\mu\nu}=0$.

The contracted Bianchi identities and the covariant conservation of the 
energy-momentum tensor give the conservation law
\begin{equation}
\nabla_\alpha H^{\mu\nu}=-\frac{\kappa^2}{2g^2}T^{\mu\nu}\nabla_\alpha g\, ,
\end{equation}
which is zero if one deals with vacuum and the coupling $g$ has a 
non-diverging value (i.e. $G_{\mu\nu}=-H_{\mu\nu}$). 
For energy conditions in ETGs, the combination of $G_{\mu\nu}$ and 
$H_{\mu\nu}$ is relevant while, in GR,
one needs only the conditions for the Einstein tensor. Specifically, the 
extended SEC has the form
\begin{equation}
g(\Psi^i)\left(R_{\mu\nu}+H_{\mu\nu}-\frac{1}{2}g_{\mu\nu}H\right)u^\alpha 
u^\beta\geq0 \, ,
\end{equation}
from which one concludes that the condition $R_{\mu\nu}u^\mu u^\nu \geq0$, valid for GR, does 
not guarantee the attractive
nature of gravity. In other words, also in the case where SEC is valid, one can obtain repulsive gravity 
in ETGs, in particular in $f(R)$ gravity, as discussed in \cite{Santos:2016vjg}.

Physical quantities which are measured by an observer are the components of 
the energy-momentum tensor
\begin{equation}
\label{emten}
T^{\alpha\beta}=\rho u^\alpha u^\beta +p h^{\alpha\beta} +\Pi^{\alpha\beta} 
+2q^{(\alpha}u^{\beta)}\, ,
\end{equation}
where $\rho=T_{\alpha\beta}u^\alpha u^\beta$ and 
$p=\frac{1}{3}T_{\alpha\beta}h^{\alpha\beta}$ are the energy-density
and the isotropic pressure, respectively. $\Pi^{\alpha\beta}=\left( 
h^{\alpha\sigma}h^{\beta\gamma}
- \frac{1}{3}h^{\alpha\beta}h^{\sigma\gamma}T_{\sigma\gamma}\right)$ is the 
anisotropic stress tensor
and $q^\alpha=u^\sigma T_{\sigma\gamma}h^{\alpha\gamma}$ denotes the current 
vector of the heat/energy flow.
The quantity
$h^\alpha_{\,\beta}=\delta^\alpha_{\,\beta}+u^\alpha u_\beta$, 
is an orthogonal projection tensor. 

 The last two components of \eqref{emten} vanish if one considers a perfect fluid medium. In that 
case it is very convenient to choose
an observer comoving with the fluid \cite{malc}; it means that the observer is 
at rest with respect to the flow of the fluid.

Any ETG can be described as an effective theory \cite{Capozziello:2011et} bringing the further 
geometric/field components on the r.h.s. of the Einstein field equations 
\begin{equation}\label{ein_eff}
G_{\alpha\beta}=\frac{\kappa^2}{2} T_{\alpha\beta}^\mathrm{eff}\, ,
\end{equation}
where $T_{\alpha\beta}^\mathrm{eff}$ is an effective energy-momentum tensor 
defines as $T_{\alpha\beta}^\mathrm{eff}
=T^{(m)}_{\alpha\beta}/g-\frac{\kappa^2}{2} H_{\alpha\beta}$. Here 
$T_{\alpha\beta}$ is the ordinary matter content and the
quantity $gH_{\alpha\beta}$
can be seen as an extra energy-momentum tensor related to scalar fields and curvature invariants 
\cite{Capozziello:2014bqa}. The terms in the tensor \eqref{emten} can be defined for any ETG as 
\begin{align}
\tilde{\rho}=&(gH_{\alpha\beta})u^\alpha u^\beta\, ,\\
3\tilde{p}=&(gH_{\alpha\beta})h^{\alpha\beta}\, ,\\
\tilde{\Pi}^{\alpha\beta}=&(gH_{\alpha\beta})\left( 
h^{\alpha\sigma}h^{\beta\gamma}
 - \frac{1}{3}h^{\alpha\beta}h^{\sigma\gamma}\right)\, ,\\
\tilde{q}^\alpha=&(gH_{\alpha\beta})u^\sigma h^{\alpha\gamma}\, ,
\end{align}
where a straightforward fluid-dynamical picture is restored. 
The above physical quantities can be measured by an observer 
$u_{(m)}^\alpha$ comoving with the perfect fluid
described by the energy-momentum
tensor $T_{\alpha\beta}$. Below, we will specify these considerations for the case of $f(R)$ assuming a Friedmann-Robertson-Walker (FRW) cosmology.


\section{Definitions of the Energy-Momentum Tensor in $f(R)$ Gravity \label{SecIV}}
The action of the $f(R)$ gravity is given by 
\cite{Nojiri:2010wj,Capozziello:2011et,Olmo:2011uz,Nojiri:2017ncd,Nojiri:2003ft}
\begin{equation}
\label{JGRG7}
S_{f(R)}= \int d^4 x \sqrt{-g} \left[ \frac{f(R)}{2\kappa^2}
+ \mathcal{L}_{(m)} \left(\Phi_i, g_{\mu\nu} \right) \right]\, .
\end{equation}
Here $\mathcal{L}_{(m)}$ is the Langrangian density of the matter and
$\Phi_i$'s express all the matter fields involved into dynamics.
By the variation with respect to the metric $g_{\mu\nu}$, we obtain the 
following equation,
\begin{equation}
\label{JGRG13}
\frac{1}{2}g_{\mu\nu} f(R) - R_{\mu\nu} f'(R) - g_{\mu\nu} \Box f'(R)
+ \nabla_\mu \nabla_\nu f'(R)
= - \frac{\kappa^2}{2}T^{(m)}_{\mu\nu}\, .
\end{equation}
Here $T^{(m)}_{ \mu\nu}$ is the energy-momentum tensor for the matters, which 
satisfies the conservation law,
\begin{equation}
\label{cons}
0 = \nabla^\mu T^{(m)}_{\mu\nu}=0 \, .
\end{equation}
Eq.~(\ref{JGRG13}) can be rewritten as
\begin{equation}
\label{TTT1}
R_{\mu\nu} - \frac{1}{2} g_{\mu\nu} R
= \frac{\kappa^2}{2} T_{\mu\nu} \, , \quad
T_{\mu\nu} \equiv \frac{1}{ f'(R)} \left[ T^{(m)}_{\mu\nu}
 - \frac{1}{2} \left( f'(R) R - f(R) + 2 \Box f'(R) \right) +\nabla_\mu 
\nabla_\nu f'(R) \right] \, .
\end{equation}
Then the Bianchi identity tells that $T_{\mu\nu}$ is also conserved,
\begin{equation}
\label{consT}
0 = \nabla^\mu T_{\mu\nu}=0 \, ,
\end{equation}
and $T_{\mu\nu}$ can be regarded as an effective energy-momentum tensor.
We may identify the matter independent part of $T_{\mu\nu}$ in (\ref{TTT1}) as 
a contribution from the dark energy,
\begin{equation}
\label{TTT2}
T_{\mu\nu}^{\mathrm{DE1}} \equiv \frac{1}{ f'(R)} \left[
 - \frac{1}{2} \left( f'(R) R - f(R) + 2 \Box f'(R) \right) +\nabla_\mu 
\nabla_\nu f'(R) \right] \, ,
\end{equation}
although $T_{\mu\nu}^{\mathrm{DE1}}$ is not conserved.
In general, the linear combination of $T_{ (m)\, \mu\nu}$ and
$T_{\mu\nu}$ is conserved.
Especially we may define an conserved energy-momentum tensor,
\begin{equation}
\label{TTT3}
T_{\mu\nu}^{\mathrm{DE2}} \equiv
\left( \frac{1}{ f'(R)} - 1 \right) T^{(m)}_{\mu\nu} + \frac{1}{ f'(R)} \left[
 - \frac{1}{2} \left( f'(R) R - f(R) + 2 \Box f'(R) \right) +\nabla_\mu 
\nabla_\nu f'(R) \right] \, ,
\end{equation}
which vanishes in the limit of the Einstein gravity, where $ f(R) \to R$.
Therefore we may regard $T_{\mu\nu}^{\mathrm{DE2}}$ with the conserved
energy-momentum tensor of the dark energy including the corrections from 
matters.
We should note that Eq.~(\ref{TTT1}) or (\ref{TTT3}) tells the matters 
including the dark matter interact with the dark energy.

One can also rewrite $f(R)$ gravity in the scalar-tensor form 
\cite{Nojiri:2003ft}.
By introducing the auxiliary field $A$, the action (\ref{JGRG7}) of
the $f(R)$ gravity is rewritten in the following form:
\begin{equation}
\label{JGRG21}
S=\frac{1}{2\kappa^2}\int d^4 x \sqrt{-g} \left\{ f'(A)\left(R-A\right)
+ f(A) + 2\kappa^2 \mathcal{L}_{(m)} \left(\Phi_i, g_{\mu\nu} \right)\right\}
\, .
\end{equation}
By the variation of $A$, one obtains $A=R$. Substituting $A=R$ into
the action (\ref{JGRG21}), one can reproduce the action in (\ref{JGRG7}).
Furthermore, we rescale the metric in the following way,
\begin{equation}
\label{JGRG22}
g_{\mu\nu}= \e^\sigma \tilde g_{\mu\nu}\, ,\quad \sigma = -\ln f'(A)\, .
\end{equation}
Then we obtain the Einstein frame action,
\begin{align}
\label{JGRG23}
S_E =& \frac{1}{2\kappa^2}\int d^4 x \sqrt{- \tilde g}
\left[ \tilde R - \frac{3}{2} \tilde g^{\rho\sigma}
\partial_\rho \sigma \partial_\sigma \sigma - V(\sigma)
+ 2\kappa^2 \mathcal{L}_{(m)} \left(\Phi_i,
\e^\sigma \tilde g_{\mu\nu}\right)\right] \, ,\nn
V(\sigma) =& \e^\sigma h\left(\e^{-\sigma}\right)
 - \e^{2\sigma} f\left(h\left(\e^{-\sigma}\right)\right) = \frac{A}{ f'(A)}
 - \frac{ f(A)}{ f'(A)^2}\, .
\end{align}
Here $\tilde R$ is the scalar curvature given by $\tilde g_{\mu\nu}$ and
$h\left(\e^{-\sigma}\right)$ is given by solving the equation
$\sigma = -\ln\left( 1 + f'(A)\right)=- \ln f'(A)$ as
$A=h\left(\e^{-\sigma}\right)$.
Due to the scale transformation (\ref{JGRG22}), a coupling
of the scalar field $\sigma$ with usual matter arises.

By the variation of the action (\ref{JGRG23}) with respect to the metric 
$\tilde g_{\mu\nu}$, we
obtain the Einstein equation,
\begin{equation}
\label{TT6}
\tilde R_{\mu\nu} - \frac{1}{2} \tilde g_{\mu\nu} \tilde R
= \kappa^2 \left\{ \kappa^2 \left[ \frac{3}{2} \partial_\mu \sigma \partial_\nu 
\sigma
- \frac{1}{2} \tilde g_{\mu\nu} \left( \frac{3}{2} \tilde g^{\rho\sigma}
\partial_\rho \sigma \partial_\sigma \sigma + V(\sigma) \right) \right]
+ \e^{-\sigma} T_{\mu\nu} \right\} \, .
\end{equation}
Here $T_{\mu\nu}$ is defined by
\begin{align}
\label{TT7}
T_{\mu\nu} \equiv& \left. \frac{2}{\sqrt{-g}} g_{\mu\rho} g_{\nu\sigma}
\frac{\delta}{\delta g_{\rho\sigma}} \left( \int d^4 x \sqrt{-g}
\mathcal{L}_{(m)} \left(\Phi_i, g_{\mu\nu} \right) \right)
\right|_{g_{\mu\nu}= \e^\sigma \tilde g_{\mu\nu}} \nn
=& \frac{2\e^\sigma}{\sqrt{-\tilde g}} \tilde g_{\mu\rho} \tilde g_{\nu\sigma}
\frac{\delta}{\delta \tilde g_{\rho\sigma}} \left( \int d^4 x \sqrt{-\tilde g}
\mathcal{L}_{(m)} \left(\Phi_i, \e^\sigma \tilde g_{\mu\nu} \right) \right) 
\, ,
\end{align}
and identical with $T_{\mu\nu}$ in (\ref{Turhop2}).
The Bianchi identity $\tilde \nabla^\mu \left( \tilde R_{\mu\nu}
 - \frac{1}{2} \tilde g_{\mu\nu} \tilde R \right)$ tells that the quantity
\begin{equation}
\label{TT8}
\hat T_{\mu\nu} \equiv \kappa^2 \left[ \frac{3}{2} \partial_\mu \sigma 
\partial_\nu \sigma
 - \frac{1}{2} \tilde g_{\mu\nu} \left( \frac{3}{2} \tilde g^{\rho\sigma}
\partial_\rho \sigma \partial_\sigma \sigma + V(\sigma) \right) \right]
+ \e^{-\sigma} T_{\mu\nu} \, ,
\end{equation}
has to be conserved $\tilde\nabla^\mu \hat T_{\mu\nu} = 0$.

\section{Energy-Momentum Tensor in FRW Space-time and Energy Conditions 
\label{SecV}}

By the scale transformation $g_{\mu\nu}= \e^{\sigma(t)} \tilde g_{\mu\nu}$,
the metric of the FRW universe is transformed by
\begin{equation}
\label{scale1}
ds^2 = - dt^2 + a(t)^2 \sum_{i=1}^3 \left( dx^i \right)^2 \ \to \
d{\tilde s}^2 = \e^{-\sigma(t)} ds^2 =
\e^{-\sigma(t)} \left(- dt^2 + a(t)^2 \sum_{i=1}^3 \left( dx^i \right)^2 
\right) \, .
\end{equation}
Then we define the new time coordinate $\tilde t$ and the new scale factor 
$\tilde a \left( \tilde t \right)$ by
\begin{equation}
\label{scale2}
d\tilde t \equiv \e^{-\frac{\sigma(t)}{2}} dt \, , \quad
\tilde a \left( \tilde t \right) = \e^{- \frac{\sigma\left(t\left(\tilde t 
\right) \right)}{2}}
a \left( t \left( \tilde t \right) \right) \, .
\end{equation}
We may assume the energy density $\rho$ and the pressure $p$ in the original 
frame satisfy the
equation of state with the equation of state (EoS) parameter $w$, $p=w \rho$.
We also assume they satisfy the conservation law,
\begin{equation}
\label{scale3}
0 = \frac{d\rho}{dt} + 3 H \left( \rho + p \right)
= \frac{d\rho}{dt} + 3 H \left( 1 + w \right) \rho \, .
\end{equation}
Here $H=\frac{1}{a} \frac{da}{dt}$.
Eq.~(\ref{scale3}) tells that $\rho$ behaves as
\begin{equation}
\label{scale3b}
\rho = \rho_0 a^{-3 (1+w)}\, .
\end{equation}
If we define
\begin{equation}
\label{scale4}
\tilde\rho \equiv \e^{\frac{3 \left(1 + w \right)\sigma}{2}} \rho\, , \quad
\tilde p \equiv \e^{\frac{3 \left(1 + w \right)\sigma}{2}} p \, ,
\end{equation}
we find
\begin{equation}
0 = \frac{d\tilde\rho}{d\tilde t} + 3 \tilde H \left (\tilde \rho + \tilde p 
\right) \, , \quad
\tilde p = w \tilde \rho \, .
\end{equation}
Here $\tilde H=\frac{1}{\tilde a} \frac{d\tilde a}{d\tilde t}$.
Here we should note
$\frac{d\tilde a}{d\tilde t}=\e^{\frac{\sigma(t)}{2}} \frac{d\tilde a}{d t}$.

Then if $\rho$ and $p$ satisfy any of the following energy condition in the FRW 
universe,
\begin{eqnarray}
\label{phtm11}
&\circ &\ \mbox{NEC:} \ \rho + p \geq 0\\
\label{phtm8}
&\circ &\ \mbox{WEC:} \ \rho\geq 0 \ \mbox{and}\ \rho 
+ p \geq 0 \\
\label{phtm9}
&\circ &\ \mbox{SEC:} \ \rho + 3 p \geq 0\ 
\mbox{and}\
\rho + p \geq 0\\
\label{phtm10}
&\circ &\ \mbox{DEC:}\ \rho\geq 0 \ \mbox{and}\ 
\rho \pm p \geq 0
\end{eqnarray}
$\tilde\rho$ and $\tilde p$ satisfy the same energy condition,
\begin{eqnarray}
\label{phtm11B}
&\circ &\ \mbox{NEC:} \ \tilde\rho + \tilde p \geq 0\\
\label{phtm8B}
&\circ &\ \mbox{WEC:} \ \tilde \rho\geq 0 \ \mbox{and}\
\tilde\rho + \tilde p \geq 0 \\
\label{phtm9B}
&\circ &\ \mbox{SEC:} \ \tilde \rho + 3 \tilde p \geq 
0\ \mbox{and}\
\tilde\rho + \tilde p \geq 0\\
\label{phtm10B}
&\circ &\ \mbox{DEC:}\ \tilde\rho\geq 0 \ 
\mbox{and}\
\tilde\rho \pm \tilde p \geq 0
\end{eqnarray}
For the matters with a constant EoS parameter $w$, when we assume $\rho>0$, if 
$w>-1$, the NEC (\ref{phtm11}) and the WEC (\ref{phtm8}) are satisfied.
If $\left| w \right| \leq 1$, the DEC (\ref{phtm10}) is satisfied.
On the other hand, the SEC (\ref{phtm9}) requires 
$w \geq - \frac{1}{3}$.

We now try to write the EoS parameter $w$ in a covariant form.
Because
\begin{equation}
\label{cow1}
T^{ (m)}\equiv g^{\mu\nu} T^{ (m)}_{\mu\nu}
= - \rho + 3 p = \left( -1 + 3w \right) \rho \, , \quad
T^{(m)}_{\mu\nu} T^{ (m)\mu\nu} = \rho^2 + 3 p^2 = \left( 1 + 3 w^2 \right) 
\rho^2 \, ,
\end{equation}
we obtain
\begin{equation}
\label{cow2}
\alpha \equiv \frac{T^{ (m)}_{\mu\nu}
T^{(m)\mu\nu }}{T^{ (m) 2}}
= \frac{ 1 + 3 w^2}{\left( - 1 + 3 w \right)^2} \, .
\end{equation}
By solving (\ref{cow2}) with respect to $w$, we obtain
\begin{equation}
\label{cow3}
w = \frac{3 \pm \sqrt{ 9 - \left( 9\alpha - 3 \right)\left( \alpha - 1 
\right)}} {9\alpha - 3 } \, .
\end{equation}

In the covariant form, the energy-momentum tensor can be expressed by using 
$\rho$ and $p$, as follows,
\begin{equation}
\label{Turhop}
T^{ (m)}_{\mu\nu} = \left( \rho + p \right) u_\mu u_\nu + g_{\mu\nu} p \, .
\end{equation}
Here $u_\mu$ is the four velocity of the fluid satisfying $u_\mu u^\mu = - 1$.
Then the conservation law $0=\nabla^\mu T_{\mu\nu}=0$ can be rewritten as
\begin{equation}
\label{Turhop0A}
0 = u_\mu u_\nu \partial^\mu \rho + \left(u_\mu u_\nu + g_{\mu\nu} \right) 
\partial^\mu p
+ \left( \rho + p \right) \left( \nabla^\mu u_\mu u_\nu + u_\mu \nabla^\mu 
u_\nu \right) \, .
\end{equation}
If the EoS parameter $w$ is a constant, Eq.~(\ref{Turhop}) can be rewritten as
\begin{equation}
\label{Turhop2}
T^{(m)}_{\mu\nu} = \left\{\left( 1+w\right) u_\mu u_\nu
+ w g_{\mu\nu}\right\} \rho \, .
\end{equation}
Then the conservation law in (\ref{Turhop0A}) has the following form
\begin{equation}
\label{Turhop0B}
0 = \left\{ \left( 1 + w \right) u_\mu u_\nu + w g_{\mu\nu} \right\} 
\partial^\mu \rho
+ \left( 1 + w \right) \rho \left( \nabla^\mu u_\mu u_\nu + u_\mu \nabla^\mu 
u_\nu \right) \, .
\end{equation}
By multiplying (\ref{Turhop0A}) and (\ref{Turhop0B}) with $u^\nu$, we obtain 
the reduced version of conservation,
\begin{equation}
\label{TT1}
0 = u^\mu \partial_\mu \rho + \nabla^\mu u_\mu \left(\rho + p \right)
= u^\mu \partial_\mu \rho + \left(1 + w \right) \nabla^\mu u_\mu \rho \, .
\end{equation}

Under the scale transformation, $u_\mu$ transforms as
\begin{equation}
\label{Turhop3}
u_\mu = \e^{\frac{\sigma}{2}} \tilde u_\mu\, .
\end{equation}
We also note that the connection is transformed as
\begin{equation}
\label{Turhop4}
\Gamma^\mu_{\nu\rho} = \tilde\Gamma^\mu_{\nu\rho}
 - \frac{1}{2} \left( \delta^\mu_{\ \nu} \partial_\rho \sigma
+ \delta^\mu_{\ \rho} \partial_\nu \sigma
 - \tilde g_{\nu\rho} \tilde g^{\mu\sigma} \partial_\sigma \sigma \right) \, .
\end{equation}
Then by using (\ref{scale4}), we find the reduced conservation law (\ref{TT1}), 
again,
\begin{equation}
\label{TT2}
0 = \tilde u^\mu \partial_\mu \tilde \rho
+ \left(1 + w \right) \tilde g^{\mu\nu} \tilde \nabla_\mu \tilde u_\nu
\tilde \rho \, .
\end{equation}
Here $\tilde\nabla_\mu$ is given in terms of $\tilde\Gamma^\mu_{\nu\rho}$ in 
(\ref{Turhop4}).
We should note that we did not redefine the coordinate not as in (\ref{scale2}).

Motivated in (\ref{JGRG22}), (\ref{scale4}), (\ref{Turhop2}), and 
(\ref{Turhop3}), we may define
\begin{equation}
\label{TT3}
\tilde T^{(m)}_{\mu\nu} = \e^{\frac{\left( 1 + 3w \right)\sigma}{2}}
T^{ (m)}_{\mu\nu} \, .
\end{equation}
We should note that $\tilde T^{(m)}_{\mu\nu}$ satisfies the identical energy 
conditions (\ref{phtm11B}), (\ref{phtm8B}), (\ref{phtm9B}), and (\ref{phtm10B})
with those (\ref{phtm11}), (\ref{phtm8}), (\ref{phtm9}), and (\ref{phtm10})
for $T_{\mu\nu}$.
Then we find
\begin{align}
\label{Turhop5}
\tilde\nabla^\mu \tilde T^{(m)}_{\mu\nu}
=& \tilde g^{\mu\rho} \tilde\nabla_\mu \tilde T^{(m)}_{\rho\nu} \nn
\equiv& \e^{\frac{3 \left( 1 + w \right)\sigma}{2}}
g^{\mu\rho} \left( \nabla_\mu T_{ (m)\rho\nu}
+ \frac{\left( 1 + 3w \right)}{2}\partial_\mu \sigma T^{ (m)}_{\rho\nu} \right. 
\nn
& \left. + \frac{1}{2} \left\{ \left( \delta^\tau_{\ \mu} \partial_\rho \sigma
+ \delta^\tau_{\ \rho} \partial_\mu \sigma
 - g_{\mu\rho} g^{\tau\sigma} \partial_\sigma \sigma \right) T^{ (m)}_{\tau\nu}
+ \left( \delta^\tau_{\ \mu} \partial_\nu \sigma
+ \delta^\tau_{\ \nu} \partial_\mu \sigma
 - g_{\mu\nu} g^{\tau\sigma} \partial_\sigma \sigma \right) T^{ (m)}_{\rho\tau}
\right\} \right) \nn
=& \e^{\frac{3 \left( 1 + w \right)\sigma}{2}}
g^{\mu\rho} \left( \nabla_\mu T^{ (m)}_{\rho\nu}
+ \frac{\left( - 1 + 3 w \right)}{2} \partial_\rho \sigma T^{ (m)}_{\mu\nu}
+ \frac{1}{2} \partial_\nu \sigma T^{ (m)}_{\mu\rho}\right) \, .
\end{align}
Therefore even if $T^{ (m)}_{\mu\nu}$ is conserved, that is,
$0=\nabla^\mu T^{ (m)}_{\mu\nu}$
in the Jordan frame, $\tilde T^{(m)}_{\mu\nu}$ does not conserved.
By using (\ref{JGRG22}), (\ref{scale4}), (\ref{Turhop2}), and (\ref{Turhop3}),
we find
\begin{equation}
\label{TT4}
\e^{\frac{3 \left( 1 + w \right)\sigma}{2}}
g^{\mu\rho} \left( \frac{\left( - 1 + 3 w \right)}{2} \partial_\rho \sigma
T^{ (m)}_{\mu\nu}
+ \frac{1}{2} \partial_\nu \sigma T^{ (m)}_{\mu\rho}\right)
= \frac{ \left( -1 + 3 w \right) \left( 1 + w \right) }{2}\left\{
\tilde g^{\mu\rho} \partial_\rho \sigma \tilde u_\mu \tilde u_\nu
+ \partial_\nu \sigma \right\} \tilde \rho \, .
\end{equation}
By multiplying $g^{\nu\rho}\tilde u_\rho$ with Eq.~(\ref{TT4}), we find
\begin{equation}
\label{TT5}
\e^{\frac{3 \left( 1 + w \right)\sigma}{2}}
g^{\mu\rho} \left( \frac{\left( - 1 + 3 w \right)}{2} \partial_\rho \sigma
T^{ (m)}_{\mu\nu}
+ \frac{1}{2} \partial_\nu \sigma T^{ (m)}_{\mu\rho}
\right) g^{\nu\rho}\tilde u_\rho = 0 \, ,
\end{equation}
which is consistent with (\ref{TT2}).
Eq.~(\ref{TT4}) tells that $\tilde T^{(m)}_{\mu\nu}$ conserved in case of the 
radiation $\left(w = \frac{1}{3} \right)$, which is scale invariant, and also in case of 
the cosmological constant, where $\rho$ and $p$ are invariant under the scale 
transformation as clear from (\ref{scale4}).

We now consider the energy conditions of $\hat T_{\mu\nu}$ in (\ref{TT8}).
Although $\e^{-\sigma} T^{ (m)}_{\mu\nu}$ in the r.h.s. of (\ref{TT8}) 
satisfies the identical energy conditions with $T^{ (m)}_{\mu\nu}$,
due to the contribution from $\sigma$, $\hat T_{\mu\nu}$ in (\ref{TT8}) does 
not always satisfy 
the identical energy conditions with those of $T^{ (m)}_{\mu\nu}$
(\ref{phtm11}), (\ref{phtm8}), (\ref{phtm9}), and (\ref{phtm10}).
Let write the energy density and the pressure given by $\hat T_{\mu\nu}$ as 
$\hat \rho$ and $\hat p$.
For the FRW space-time in (\ref{scale1}), they are explicitly given by
\begin{equation}
\label{TT9}
\hat\rho = \kappa^2 \left( \frac{3}{2}{\dot\sigma}^2 + V(\sigma) \right)
+ \e^\sigma \rho \, , \quad
\hat p = \kappa^2 \left( \frac{3}{2}{\dot\sigma}^2 - V(\sigma) \right)
+ \e^\sigma p \, .
\end{equation}
We should note that the potential $V(\sigma)$ can be negative in general.
Eqs.~(\ref{scale1}) tells that even by the scale transformation (\ref{JGRG22}), 
the FRW space-time
is transformed into the FRW space-time, therefore, where the space-time is 
expanding or shrinking,
and therefore the energy density $\hat\rho$ should be positive,
\begin{equation}
\label{TT9B}
\hat\rho>0 \, ,
\end{equation}
which gives the lower bound for $V(\sigma)$,
\begin{equation}
\label{TT10}
V(\sigma) > - \frac{3\kappa^2 }{2}{\dot\sigma}^2 - \e^\sigma \rho
> - \e^\sigma \rho\, .
\end{equation}
Because now we have
\begin{equation}
\label{TT11}
\hat\rho + \hat p = 3 \kappa^2 {\dot\sigma}^2 
+ \e^\sigma \left( \rho + p \right)
\geq \e^\sigma \left( \rho + p \right) \, ,
\end{equation}
Therefore if $\rho$ and $p$ satisfy the NEC (\ref{phtm11}) and WEC
(\ref{phtm8}), $\hat\rho$ and $p$ also satisfy the Conditions.
We also find
\begin{equation}
\label{TT12}
\hat\rho + 3 \hat p
= 2 \kappa^2 \left( 3 {\dot\sigma}^2 - V(\sigma) \right) + \rho + 3 p \, ,
\end{equation}
which tells that when $\rho$ and $p$ satisfy the SEC 
(\ref{phtm9}),
$\hat\rho$ and $\hat p$ also satisfy the SEC if
$3 {\dot\sigma}^2 > V(\sigma)$.
On the other hand, because
\begin{equation}
\label{TT13}
\hat\rho - \hat p = 2 \kappa^2 V(\sigma) + \e^\sigma \left( \rho - p \right) 
\, ,
\end{equation}
if $\rho$ and $p$ satisfy the DEC (\ref{phtm10}) and the potential $V(\sigma)$
is positive, $\hat\rho$ and $\hat p$ also satisfy the Dominant Energy Condition.

\section{The case of power-law $f(R)$ gravity \label{SecVI}}

Let us now assume that $f(R)$ behaves as $ f(R) \propto f_0 R^m$.
When we include the contributions from the matter with a
constant EoS parameter $w$, if we assume the FRW universe (\ref{scale1}),
the solution is given by
\begin{align}
\label{M8}
& a=a_0 t^{h_0} \, ,\quad h_0\equiv \frac{2m}{3(1+w)} \, ,\nn
& a_0\equiv \left[-\frac{3f_0h_0}{\kappa^2 \rho_0}\left(-6h_0 + 12
h_0^2\right)^{m-1}
\left\{\left(1-2m\right)\left(1-m\right)
 - (2-m)h_0\right\}\right]^{-\frac{1}{3(1+w)}}\, .
\end{align}
Here $\rho_0$ is defined in (\ref{scale3b}).
Then the effective EoS parameter, which is given by $T_{\mu\nu}$ in
(\ref{TTT1}), is
\begin{equation}
\label{JGRG20}
w_\mathrm{eff}= -1 + \frac{w+1}{m}\, .
\end{equation}
Then even if $w>-1$, when $m<0$, all the energy conditions (\ref{phtm11}), 
(\ref{phtm8}),
(\ref{phtm9}), and (\ref{phtm10}) are not satisfied for $T_{\mu\nu}$.
Because $R$ behaves as $A=R\propto \frac{1}{t^2}$, $\e^{-\sigma} = f'(A)$ 
behaves as
$\e^{-\sigma} \propto t^{- 2\left(m-1\right)}$.
Then the equations in (\ref{scale2}) show that
\begin{equation}
\label{TT14}
\tilde t \propto t^{2 - m}\, , \quad
\tilde a \propto t^{- \left(m-1\right) + \frac{2m}{3(1+w)}}
= t^{\frac{- 3 (m - 1 ) w - m + 3}{3(1+w)}}
\propto {\tilde t}^{- \frac{3 (m - 1 ) w - m + 3}{3(2 - m)(1+w)}} \, .
\end{equation}
In case that the matter with the EoS parameter $w$ minimally couples with 
gravity,
the scale factor behaves as $a \propto t^{\frac{2}{3(1+w)}}$, which shows 
that the effective
EoS parameter $w^\mathrm{E}_\mathrm{eff}$ in the Einstein frame, which is 
defined by using
$\hat T_{\mu\nu}$ in (\ref{TT8}), is given by
\begin{equation}
\label{TT15}
w^\mathrm{E}_\mathrm{eff} = - 1 + \frac{2(2 - m)(1+w)}{-3 (m - 1 ) w - m + 3}
= - 1 + \frac{2(2 - m)(1+w)}{- (3w + 1)m + 3(1 + w)} \, .
\end{equation}
Then even if $w>-\frac{1}{3}$, when
\begin{equation}
\label{TT16}
\frac{3(w+1)}{3w + 1} = 1 +\frac{2}{3w + 1} < m < 2 \, ,
\end{equation}
we find $w^\mathrm{E}_\mathrm{eff} < -1$, all the energy conditions 
(\ref{phtm11}),
(\ref{phtm8}), (\ref{phtm9}), and (\ref{phtm10}) are not satisfied but the 
inequality
(\ref{TT16}) is consistent when $w>\frac{1}{3}$, which might be unnatural.

Eq.~(\ref{M8}) or (\ref{JGRG20}) tells that there occurs the Big Rip 
singularity in the
Jordan frame when $\frac{w+1}{m}<0$ and therefore all the energy conditions
(\ref{phtm11}), (\ref{phtm8}),
(\ref{phtm9}), and (\ref{phtm10}) are not satisfied for $T_{\mu\nu}$.
We should note that the energy conditions of the matter follow the relations 
which have been mentioned after (\ref{phtm10B}):
\begin{itemize}
\item When $w>-1$, the NEC (\ref{phtm11})
and the WEC (\ref{phtm8}) are satisfied.
\item When $\left| w \right| \leq 1$, the DEC 
(\ref{phtm10}) is satisfied.
\item When $w \geq - \frac{1}{3}$, the SEC (\ref{phtm9}) is 
satisfied.
\end{itemize}

In the Einstein frame, if we define the energy-momentum tensor of the matter
by (\ref{TT3}), all the energy conditions of the matter do not change from 
those in the Jordan frame.
For the total energy-momentum $\hat T_{\mu\nu}$ (\ref{TT8}) 
in the Einstein frame, by using (\ref{TT15}), we find
\begin{itemize}
\item When $\frac{2(2 - m)(1+w)}{- (3w + 1)m + 3(1 + w)}>0$,
the NEC (\ref{phtm11}) and the WEC (\ref{phtm8}) are satisfied. 
That is,
\begin{itemize}
\item If $w<-1$, $m<\frac{3(w+1)}{3w + 1}$ or $m>2$.
\item If $-1<w< - \frac{1}{3}$, $\frac{3(w+1)}{3w + 1}<m<2$.
\item If $- \frac{1}{3} < w < \frac{1}{3}$, $m<2$ or $m>\frac{3(w+1)}{3w + 1}$.
\item If $w> \frac{1}{3}$, $m<\frac{3(w+1)}{3w + 1}$ or $m>2$.
\end{itemize}
or
\begin{itemize}
\item If $m<0$, $w<-1$ or $w> - \frac{3-m}{3\left( 1 - m \right)}$. 
\item If $0<m<1$, $w< - \frac{3-m}{3\left( 1 - m \right)}$ or $w>-1$. 
\item If $1<m<2$, $-1<w - \frac{3-m}{3\left( 1 - m \right)}$.
\item If $m>2$, $w<-1$ or $w> - \frac{3-m}{3\left( 1 - m \right)}$. 
\end{itemize}

\item When $2\geq \frac{2(2 - m)(1+w)}{- (3w + 1)m + 3(1 + w)}\geq 0$,
the DEC (\ref{phtm10}) is satisfied. That is,
\begin{itemize}
\item If $w<-1$, $2<m<\frac{1+w}{2w}$ or $m>2$.
\item If $-1<w< - \frac{1}{3}$, $\frac{1+w}{2w}<m<2$.
\item If $- \frac{1}{3} < w < 0$, $\frac{1+w}{2w}<m<2$.
\item If $0 < w < \frac{1}{3}$, $m<2$ or $m>\frac{1+w}{2w}$.
\item If $w> \frac{1}{3}$, $m<\frac{1+w}{2w}$ or $m>2$.
\end{itemize}
or
\begin{itemize}
\item $m<0$, $w<-1$ or $w>\frac{1}{2m-1} $. 
\item $0<m<\frac{1}{2}$, $w>-1$ or $w<\frac{1}{2m-1} $. 
\item $\frac{1}{2}<m<1$, $-1<w< \frac{1}{2m-1} $. 
\item $1<m<2$, $-1<w<\frac{1}{2m-1}$. 
\item $m>2$, $w<-1$ or $w>\frac{1}{2m-1}$. 
\end{itemize}
\item When $\frac{2(2 - m)(1+w)}{- (3w + 1)m + 3(1 + w)} \geq \frac{2}{3}$,
the SEC (\ref{phtm9}) is satisfied. That is,
\begin{itemize}
\item If $w<-1$, $m<\frac{3(1+w)}{2}$ or $m>2$.
\item If $-1<w< - \frac{1}{3}$, $\frac{3(w+1)}{3w + 1}<m<\frac{3(1+w)}{2}$.
\item If $- \frac{1}{3} < w < \frac{1}{3}$, $m<\frac{3(1+w)}{2}$
or $m>\frac{3(w+1)}{3w + 1}$.
\item If $w> \frac{1}{3}$, $m<\frac{3(w+1)}{3w + 1}$ or $m>\frac{3(1+w)}{2}$.
\end{itemize}
or
\begin{itemize}
\item If $m<0$, $w<-1 + \frac{2}{3}m$ or $w> - \frac{3-m}{3\left(1-m\right)}$.
\item If $0<m<1$, $w< - \frac{3-m}{3\left(1-m\right)}$ or $w>-1 + \frac{2}{3}m$. 
\item If $1<m<2$, $-1 + \frac{2}{3}m < w < - \frac{3-m}{3\left(1-m\right)}$. 
\item If $m>2$, $- \frac{3-m}{3\left(1-m\right)} < w < -1 + \frac{2}{3}m$.
\end{itemize}
\end{itemize}
We should note that in the Einstein frame, when 
$\frac{2(2 - m)(1+w)}{- (3w + 1)m + 3(1 + w)}<0$, 
we have the phantom phase when 
$0<\frac{2(2 - m)(1+w)}{- (3w + 1)m + 3(1 + w)}<\frac{2}{3}$, 
we have quintessence phase, and when 
$\frac{2(2 - m)(1+w)}{- (3w + 1)m + 3(1 + w)}>\frac{2}{3}$, 
we have the deceleratedly expanding universe. 

The above results are summarized in TABLE~\ref{Table1} and TABLE~\ref{Table2}.

\begin{table}
\caption{The region satisfying the energy conditions when we vary $w$. 
\label{Table1}}
\begin{tabular}{|c||c|c|c|} \hline \hline
& NEC, WEC & DEC & SEC \\ 
\hline \hline
$w<-1$ & $m<\frac{3(w+1)}{3w + 1}$ or $m>2$ & 
$2<m<\frac{1+w}{2w}$ or $m>2$ &
$\frac{3(1+w)}{2}$ or $m>2$ \\
\hline
$-1<w<-\frac{1}{3}$ & $\frac{3(w+1)}{3w + 1}<m<2$ & 
$\frac{1+w}{2w}<m<2$ & 
$\frac{3(w+1)}{3w + 1}<m<\frac{3(1+w)}{2}$ \\ 
\hline
$- \frac{1}{3} < w < 0$ & $m<2$ or $m>\frac{3(w+1)}{3w + 1}$ &
$\frac{1+w}{2w}<m<2$ &
$m<\frac{3(1+w)}{2}$ or $m>\frac{3(w+1)}{3w + 1}$ \\
\hline
$0 < w < \frac{1}{3}$ & $m<2$ or $m>\frac{3(w+1)}{3w + 1}$ &
$m<2$ or $m>\frac{1+w}{2w}$ &
$m<\frac{3(1+w)}{2}$ or $m>\frac{3(w+1)}{3w + 1}$ \\
\hline
$w>\frac{1}{3}$ & $m<\frac{3(w+1)}{3w + 1}$ or $m>2$ & 
$m<\frac{1+w}{2w}$ or $m>2$ & 
$m<\frac{3(w+1)}{3w + 1}$ or $m>\frac{3(1+w)}{2}$ \\
\hline \hline 
\end{tabular}
\end{table}

\begin{table}
\caption{The region satisfying the energy conditions when we vary $m$. 
\label{Table2}}
\begin{tabular}{|c||c|c|c|} \hline \hline
& NEC, WEC & DEC & SEC \\ 
\hline \hline
$m<0$ & $w<-1$ or $w> - \frac{3-m}{3\left( 1 - m \right)}$ 
& $w<-1$ or $w>\frac{1}{2m-1} $ 
& $w<-1 + \frac{2}{3}m$ or $w> - \frac{3-m}{3\left(1-m\right)}$ \\
\hline
$0<m<\frac{1}{2}$ & $w< - \frac{3-m}{3\left( 1 - m \right)}$ or $w>-1$
& $w>-1$ or $w<\frac{1}{2m-1} $ 
& $w< - \frac{3-m}{3\left(1-m\right)}$ or $w>-1 + \frac{2}{3}m$ \\ 
\hline
$\frac{1}{2} < m < 1$ & $w< - \frac{3-m}{3\left( 1 - m \right)}$ or $w>-1$
& $-1<w< \frac{1}{2m-1} $ 
& $w< - \frac{3-m}{3\left(1-m\right)}$ or $w>-1 + \frac{2}{3}m$ \\
\hline
$1 < m < 2$ & $-1<w - \frac{3-m}{3\left( 1 - m \right)}$
& $-1<w<\frac{1}{2m-1}$ 
& $-1 + \frac{2}{3}m < w < - \frac{3-m}{3\left(1-m\right)}$ \\
\hline
$m>2$ & $w<-1$ or $w> - \frac{3-m}{3\left( 1 - m \right)}$ 
& $w<-1$ or $w>\frac{1}{2m-1}$ 
& $- \frac{3-m}{3\left(1-m\right)} < w < -1 + \frac{2}{3}m$ \\
\hline \hline 
\end{tabular}
\end{table}

We have considered the case that $f(R)$ behaves as 
$f(R) \propto f_0 R^m$, which may be realized in some limit or any 
extremal circumtance 
as in the early universe like inflation. 
Not in such an extremal case, if $f(R)$ is a smooth function of 
$R$, for example, $f(R) \sim \e^{\alpha R}$ and we consider 
the era when $R=R_0$ in the 
background, we can expand $f(R)$ around the background curvature 
$R_0$ as follows, 
\begin{equation}
f(R) = f(R_0) + f'(R_0) \left( R - R_0 \right) 
+ \mathcal{O} \left( \left( R - R_0 \right)^2 \right) \, .
\end{equation}
The terms of $\mathcal{O} \left( \left( R - R_0 \right)^2 \right) $ are 
subdominant and we may neglect them. 
Then we can identify $f(R_0) - f'(R_0) R_0$ as a cosmological constant 
and $f'(R_0)$ as the inverse of Newton's gravitational constant. 
Therefore the gravity can be described by the Einstein gravity and 
therefore the energy conditions are not changed from those in the 
Einstein gravity. 
Of course, if we include 
$\mathcal{O} \left( \left( R - R_0 \right)^2 \right) $ corrections, 
there could be small deviation of the effective energy conditions. 

In the Einstein frame, when $f(R) = f_0 R^m$, Eq.~(\ref{JGRG23}) gives 
the following potential, 
\begin{equation}
\label{JV}
V(\sigma) = \frac{m - 1}{m^2 f_0^{\frac{2m-3}{m-1}}} 
\e^{\frac{m-2}{m-1}\sigma} \, .
\end{equation}
Then if we define $\tilde\rho$ and $\tilde p$ by (\ref{scale4}), 
the energy conditions (\ref{phtm11B}), (\ref{phtm8B}), 
(\ref{phtm9B}), and (\ref{phtm10B}) for $\tilde\rho$ and $\tilde p$ 
do not changed from the energy conditions (\ref{phtm11}), 
(\ref{phtm8}), (\ref{phtm9}), and (\ref{phtm10}) 
for the original energy density $\rho$ and presure $p$. 
Therefore in the Einstein frame, as in the Einstein gravity, 
we have NEC when $w>-1$, DEC when $\left| w \right| \leq 1$, 
and SEC when $w\geq - \frac{1}{3}$. 

We should note $w_\mathrm{eff}$ given in (\ref{JGRG20}) and 
$w^\mathrm{E}_\mathrm{eff}$ given in (\ref{TT15}) are 
different from the EoS parameter $w$, which is defined by $w=\frac{p}{\rho}$ 
using the matter energy density and the pressure of the matter as given 
before (\ref{scale3}). 
The EoS parameter $w$ is defined in the Jordan frame but it does not change even 
in the Einstein frame if we use $\tilde\rho$ and $\tilde p$ in (\ref{scale4}), 
that is, $\frac{p}{\rho} = \frac{\tilde p}{\tilde\rho}=w$. 
On the other hand, $w_\mathrm{eff}$ is given by $T_{\mu\nu}$ in (\ref{TTT1}) 
and $w^\mathrm{E}_\mathrm{eff}$ is given by 
$\hat T_{\mu\nu}$ in (\ref{TT8}).

As a concrete example, we consider the case that the matter is dust with $w=0$, where 
all the energy conditions for the matter energy-momentum tensor $T^{ (m)}_{\mu\nu}$ 
(\ref{Turhop}) in the Jordan frame and $\tilde T^{(m)}_{\mu\nu}$ (\ref{TT3}) in the Einstein frame 
are satisfied. 
Eq.~(\ref{M8}) or (\ref{JGRG20}) tells that if $m$ is negative, there is a Big 
Rip singularity at $t=0$ in the Jordan frame. 
This tells the dark energy-momentum tensor $T_{\mu\nu}^{\mathrm{DE1}}$ in (\ref{TTT2}) or 
$T_{\mu\nu}^{\mathrm{DE2}}$ in (\ref{TTT3}) in the Jordan frame does not satisfy any energy 
condition. 
As clear from the above analysis, even if $m<0$, all the energy conditions can be 
satisfied for the total energy momentum tensor $\hat T_{\mu\nu}$ (\ref{TT8}) 
in the Einstein frame.
We should note that Eq.~(\ref{TT14}) tells the time for the Big Rip singularity 
in the Jordan
frame, $t=0$, corresponds to $\tilde t \to \infty$ and there does not occur the 
singularity in the finite time in the Einstein frame. See also \cite{Bahamonde:2016wmz}.

\section{Discussion and Conclusions \label{SecVII}}

The role of energy conditions is crucial to define self-consistent and physically motivated theories of gravity. 
Specifically, any fluid assumed as source of the field equations has to be compatible with 
causal and geodesic 
structure of space-time so then energy conditions can be seen as a sort of ``selection rule''
for viable relativistic theories. 
In the case of ETGs, the further degrees of freedom related to geometric invariants and scalar fields can be represented as contributions in the effective stress-energy tensor. This means that, also by preserving the physical meaning of energy conditions, dynamics can be affected and modified with respect to the standard GR because the stress-energy tensor results modified with respect to the one of standard matter.
 
In this paper, we have considered energy conditions in $f(R)$ cosmology. The main role for the discussion is played by the definition of the stress-energy tensor that can be defined in the Jordan and in the Einstein frame under a conformal transformation. The further degrees of freedom of $f(R)$ gravity can be modeled out as a scalar field that modifies the energy-matter content and the affects the dynamics. In particular, we showed that dark energy behaviors, and then generalization of EoS giving rise to accelerated (repulsive) gravity are compatible with energy conditions that, in any case, preserve causality and geodesis structure. 
Specifically, in the case of power-law $f(R)$ gravity, both the EoS (i.e. $w$) and the specific cosmological models (given by the power $m$ of $R^m$) can be combined with energy conditions. The emerging classification selects viable ranges of $w$ and $m$. In particular, the accelerating/decelerating behaviors (then dark energy behaviors), the presence/absence of Big Rip singularities and other important cosmological features, strictly depend on the given energy condition that have to be satisfied.

 In  general,   theoretical constraints on 
 the functional form of $f(R)$ can be derived from energy conditions.  As discussed in \cite{Santos:2016vjg, Capozziello:2014bqa,Capozziello:2013vna}, we can  consider the cosmic fluid evolution,  given by the Raychaudhury equation for the congruence of timelike geodesics, as the dynamical equation to be compared with the energy conditions.  It is
 \begin{equation}
\frac{d\theta}{d\tau}= - \frac{\theta^2}{3}-\sigma_{\mu\nu}\sigma^{\mu\nu}+w_{\mu\nu}w^{\mu\nu}
-R_{\mu\nu}\xi^{\mu}\xi^{\nu}\,.
\label{Raychaudhuri tempo}
\end{equation}
Here $\xi^{\mu}=dx^{\mu}/d\tau$ is a  tangent vector,  $\theta,\sigma^{\mu\nu},w^{\mu\nu}$ are  the expansion, the shear and the twist of the congruence of geodesics respectively;  $\tau$ is the proper time of an observer moving along a geodesic. According to this approach, 
it is possible to define a function $M_{\xi^{\mu}}\equiv -R_{\mu\nu}\xi^\mu\xi^\nu$ which is related to the geodesic focusing. The sign of such a function is crucial: for $M_{\xi^{\mu}}>0$,  we have geodesic defocusing; for $M_{\xi^{\mu}}<0$, there is geodesic focusing, for $M_{\xi^{\mu}}=0$ there is no contribution. For example, according to the notation in \cite{Santos:2016vjg} where $F(R)=R+f(R)$,
the SEC is satisfied if
\begin{equation}
M_{\xi^{\mu}}\leq\frac{ Rf'-f+(\nabla^{\alpha}\nabla_{\alpha} -2\xi^{\mu}\xi^{\nu}\nabla_{\mu}\nabla_{\nu})f'}{2(1+f')},
\label{M}
\end{equation}
and then the form of $f(R)$ results constrained.
Furthermore, more precise  constraints can come  from cosmography.  In fact, being the cosmographic parameters
  Hubble $H$, deceleration $q$, jerk $j$, and snap $s$ parameters, defined  as
\begin{eqnarray}
H=\frac{\dot{a}}{a},  \qquad q=-\frac{1 }{H^{2}}\frac{\ddot{a}}{a},  \qquad j= \frac{1 }{H^{3}}\frac{\dddot{a}}{a}, \qquad s=\frac{1}{H^{4}}\frac{\ddddot{a}}{a}\,,
\label{parameters}
\end{eqnarray}
 it is worth to define the Ricci scalar and its derivatives  in FRW metric as:
\begin{eqnarray}
&&R =  6H^{2}(1-q)\,,\nonumber\\
&&\dot{R} = 6H^{3}(j-q-2)\,,\nonumber\\
&&\ddot{R} =  6H^{4}(s+q^{2}+8q+6)\,.
\label{ricci scalar}
\end{eqnarray}
According to  observations
\cite{Riess:1998cb, Riess:1998dv, Perlmutter:1998np,Spergel:2003cb,Eisenstein:2005su}, the cosmographic  parameters (\ref{parameters}) can be fixed in a model independent way. Starting from Eq.(\ref{M}), we get 
\begin{equation}
M_{\xi^{\mu}_{FRW}}\leq\frac{- f/2 + c_1f' + c_2f'' + c_3f'''}{1+f'}\,,
\label{geral4time}
\end{equation}
where primes indicate derivative with respect to $R$.
The coefficients $c_i$ can be related to   cosmography as
\begin{eqnarray}
c_1 & = & 3(1-q)H^{2}\,, \nonumber \\
c_2 & = & -9(s+j + q^{2}+7q+4)H^{4}\,, \nonumber \\
c_3 & = & - 54(j-q-2)^{2}H^{6}\,.
\end{eqnarray}
Eq. (\ref{geral4time}) fixes the range of possible $f(R)$ models according to the SEC as soon as the set of numbers $\{H, q, j, s\}$ is given by the observations.

Besides,   theoretical and observational constraints can come from other physical requirements like   the absence of ghost modes, gravitational wave constraints, 
compatibility with fifth force and large scale-structure, see for example  \cite{DeMartino:2018yqf, DeLaurentis:2018ahr, DeMartino:2017ztt, deMartino:2015zsa}.

In a forthcoming paper, this approach will be developed for other ETGs and confronted to the observations.

\section*{Acknowledgments}
This work is supported (in part) by MEXT KAKENHI Grant-in-Aid for Scientific Research 
on Innovative Areas ``Cosmic Acceleration'' (No. 15H05890) (SN), by MINECO (Spain), project
FIS2016-76363-P (SDO),  by CSIC I-LINK1019 Project (SC, SN, and SDO), and by JSPSS17116 short-term fellowship
(SDO).
SC is supported by Istituto Nazionale di Fisica Nucleare (INFN).
This research was started while SN was visiting the Institute for Space Sciences, ICE-CSIC 
in Bellaterra, Spain (SN thanks Emilio Elizalde and SDO, and the rest of the members of the ICE 
for very kind hospitality).
This article is based upon work from COST Action CA15117 ``Cosmology and Astrophysics Network for Theoretical Advances and Training Actions'' (CANTATA), supported by COST (European Cooperation in Science and Technology).

\end{document}